\documentclass[12pt]{article}
\usepackage[dvips]{graphicx}
\usepackage{amssymb}
\usepackage{amsmath}
\usepackage{epstopdf}
\usepackage{color}
\usepackage{soul}
\usepackage{cite}

\makeatletter

\@addtoreset{equation}{section}
\makeatletter

\begin{document}

\begin{titlepage}
\vfill

\begin{center}
\baselineskip=16pt
{\Large\bf 
A class of rotating hairy black holes \\ in arbitrary dimensions
}
\vskip 0.5cm

{\large Cristi\'an Erices and Cristi{\'a}n Mart\'{\i}nez}

{\textit{
    Centro de Estudios Cient\'{\i}ficos (CECs), Av. Arturo Prat 514, Valdivia, Chile. }\\[0.1cm]
	\texttt{erices@cecs.cl, martinez@cecs.cl}
	 }
\vspace{6pt}
\end{center}
\vskip 0.2in
\par
\begin{center}
{\bf Abstract}
\end{center}

\begin{quote}
A class of exact rotating black hole solutions of gravity nonminimally coupled to a self-interacting scalar field in arbitrary dimensions is presented. These spacetimes are asymptotically locally anti-de Sitter manifolds and have a Ricci-flat event horizon hiding a curvature singularity at the origin. The scalar field is real and regular everywhere and its effective mass, coming from the nonminimal coupling with the scalar curvature, saturates the Breitenlohner-Freedman bound for the corresponding spacetime dimension. The rotating black hole is obtained by applying an improper coordinate transformation to the static one. Although both spacetimes are locally equivalent, they are globally different, as it is confirmed by the nonvanishing angular momentum of the rotating black hole. It is found that the mass is bounded from below by the angular momentum in agreement with the existence of an event horizon. The thermodynamical analysis is carried out in the grand canonical ensemble.  The first law is satisfied and a Smarr formula is exhibited. The thermodynamical local stability of the rotating hairy black holes is established from their Gibbs free energy. However, the global stability analysis establishes that the vacuum spacetime is always preferred over the hairy black hole. Thus, the hairy black hole is likely to decay into the vacuum one for any temperature.
\vfill
\vskip 2.mm
\end{quote}
\end{titlepage}


\tableofcontents

\newpage

\section{Introduction} \label{introduction}

In the last years a number of static black holes dressed with scalar field have been found in different dimensions. These black holes circumvent the well-known no-hair conjecture by violating some of its assumptions. For instance, by introducing a cosmological constant, considering a suitable potential or allowing non-minimal couplings. In this way, for a self-interacting real scalar field nonminimally coupled to the Ricci invariant, exact black hole solutions has been found in three \cite{MZ, HMTZ01, Xu}, four \cite{Bocharova, Bekenstein1, MTS, MDG, Charmousis, Anabalon02, Caldarelli01, Caldarelli02, Astorino01, Eloy}  and even in higher dimensions \cite{Vanzo01, Martinez_proceeding, Mokhtar01, Mokhtar02, Oliva, Lu_hairy}, including in some cases the presence of other matter fields. Additionally, other couplings to a single scalar field, such that those considered in \cite{Rinaldi, Anabalon01, Erices1}, produce black hole solutions, but in this cases the derivative of the scalar field is not continuous on the horizon, even in the case where a nonvanishing cosmological constant is considered.

Undoubtedly, four dimensional black holes endowed with angular momentum in presence of matter fields have a crucial role in astrophysics. Moreover, their three and higher dimensional counterparts are interesting objects from a theoretical point of view. However, as opposite to the static case, exact rotating hairy black holes with a conformally coupled scalar field are almost absent in the literature \cite{Natsuume, Anabalon03, Caldarelli03, Astorino02}. Thus, it is not surprising that exact rotating black holes with a self-interacting scalar field have not been reported previously for arbitrary $n$ spacetime dimensions. The lack of stationary hairy solutions comes not only from the aforementioned no-hair conjecture, but also arises from the complexity of the field equations, even in the simplest case of four-dimensional General Relativity with a minimally coupled self-interacting real scalar field.
 
In this article, a class of exact rotating black holes in arbitrary dimensions with a non-minimally coupled self-interacting scalar field is presented.  The key point for obtaining this solution lays in the fact that an improper coordinate transformation acting on a static metric could generate a stationary one with nonvanishing angular momentum. This nontrivial effect was shown by Stachel in \cite{Stachel} and the physical interpretation of the parameters of this local transformation was studied in \cite{MacCallum}.  This procedure, originally established in four dimensions, has  been also used  in three dimensions \cite{Deser}. Remarkable examples of stationary spacetimes that can be obtained from static ones, are the neutral \cite{BTZ, BTZgeometry} and electrically charged rotating  BTZ black holes \cite{BTZcharged}. In four dimensions, the method was firstly considered by Lemos \cite{Lemos01} in order to get the rotating black string and then applied on other static vacuum \cite{MacCallumSantos} and electrically charged solutions \cite{Awad, Hendi02}. However, this procedure does not always provide the general stationary solution for a given theory. This is because the system could contain stationary solutions without a static limit, as it was noted in \cite{Erices3}.
 
The article is organized as follows. Section \ref{secstatic} presents a static hairy black hole solution in $n$ spacetime dimensions. The black hole behaves asymptotically as a locally anti-de Sitter (AdS) spacetime. The event horizon is a Ricci flat $(n-2)$-surface covering a curvature singularity at the origin and the scalar field is real and regular everywhere. The mass of the scalar field saturates the Breitenlohner-Freedman  bound \cite{BF01, BF02, Townsend}, which ensures the perturbative stability of AdS spacetime, when the non-minimal coupling parameter is chosen to have a precise dependence on the dimension. 
A stationary class of hairy black hole in arbitrary dimensions is introduced in Sec.  \ref{secrotating}. These black holes are built from the previous static ones  by performing an improper transformation given by a Lorentz boost in the plane spanned by the time and a spacelike direction. In Sec.  \ref{seccharges}  the conserved charges, mass and angular momentum,  are computed using the Regge-Teitelboim procedure. It is found that the black hole mass is bounded from below by the angular momentum, which is in complete agreement with the condition for the existence of an event horizon. The thermodynamical aspects of the rotating hairy black holes are studied in Sec. \ref{thermo}. First,  the basic thermodynamical quantities are determined showing that the first law holds and then a Smarr formula is obtained. By setting the thermodynamical system in the grand canonical ensemble, the analysis of the second derivatives of the Gibbs free energy respect to the intensives variables implies the local thermal stability of the rotating  black hole in any dimension. Nevertheless, the global stability analysis shows that the vacuum configuration is always favored over the hairy one. Some final remarks are included in Sec. \ref{remarks}.

\section{Static hairy black holes} \label{secstatic}

We consider gravity with a single non-minimally coupled self-interacting
real scalar field in $n \ge 3$ spacetime  dimensions described by the action 
\begin{equation}
I[g_{\mu \nu },\Phi ] =\int d^{n}x\sqrt{-g}\left[ \frac{
R}{2 \kappa}-\frac{1}{2}g^{\mu \nu }\partial _{\mu }\Phi \partial
_{\nu }\Phi -\frac{\xi}{2}R\Phi ^{2}-V(\Phi)\right],
\label{action2nd}
\end{equation}
where $\kappa$ is the gravitational constant,  $\xi $ is the parameter describing the non-minimal coupling, and $V(\Phi)$ corresponds to a self-interaction potential. The corresponding field equations are
\begin{eqnarray}  \label{field-eqs}
G_{\mu \nu }&=&\kappa T_{\mu \nu },  \label{Einstein-eqs} \\
\square \Phi  &=&\xi R\Phi +\frac{d V(\Phi)}{d\Phi},  \label{KG} 
\end{eqnarray}
where the energy-momentum tensor is given by
\begin{equation}
T_{\mu \nu }=\partial _{\mu }\Phi \partial _{\nu }\Phi -\frac{1}{2}
g_{\mu \nu }g^{\alpha \beta }\partial _{\alpha }\Phi \partial _{\beta }\Phi +
\xi [g_{\mu \nu }\square -\nabla _{\mu }\nabla _{\nu }+G_{\mu \nu
}]\Phi ^{2}-g_{\mu \nu }V(\Phi).  \label{Tuv}
\end{equation}

Choosing the non-minimal parameter as  
\begin{equation}
\xi =\frac{n-1}{4 n},
\label{confparam}
\end{equation}
and the static ansatz 
\begin{equation} \label{staticsolution}
ds^{2}=-f(r)dt^{2}+\frac{dr^{2}}{f(r)}+r^{2}d\sigma^{2},  \quad \Phi=\Phi(r),
\end{equation}
where the line element $d\sigma ^{2}$, which does not depend on the radial coordinate $r$, denotes the metric of a $(n-2)-$dimensional Euclidean Ricci-flat manifold $\mathcal{M}_{n-2}$,   a solution of the field equations given by the lapse function  
\begin{equation} \label{f}
f(r)=\frac{r^2}{l^2}+\alpha  r^2 \log \left(1-\left(\frac{a}{r}\right)^{n-1}\right),
\end{equation}
and the scalar field
\begin{equation} \label{scalar}
\Phi(r)=\sqrt{\frac{4 n}{\kappa  (n-1)}}\left(\frac{a}{r}\right)^{\frac{n-1}{2}},
\end{equation}
is obtained. 
Here $a$ is an integration constant, the parameter $l$ is the $n$-dimensional AdS radius, and $\alpha$ is a coupling constant appearing in the self-interaction potential
\begin{align} \label{potential}
V(\Phi) = & -\frac{ (n-2) (n-1)}{2 \kappa  l^2}+\frac{\alpha  (n-1)^2 \Phi ^2 \left(-\kappa  \Phi ^2+4 n^2+n \left(\kappa  \Phi ^2-8\right)\right)}{8 n \left(n \left(\kappa  \Phi^2-4\right)-\kappa  \Phi^2\right)} \nonumber \\ 
 &-\frac{\alpha  (n-2) (n-1) \log \left(1-\frac{\kappa  (n-1) \Phi ^2}{4 n}\right)}{2 \kappa }.
\end{align}

The Ricci scalar for (\ref{staticsolution}-\ref{f}) reads
\begin{equation} \label{Rscalar}
R=(n-1)\left[-\frac{n}{l^2}+\alpha \left(\frac{a}{r}\right)^{n-1}\frac{(2n-1)(\frac{a}{r})^{n-1}-n}{(1-(\frac{a}{r})^{n-1})^2}-\alpha n \log\left(1-\left(\frac{a}{r}\right)^{n-1}\right)\right],
\end{equation}
which determines the existence of curvature singularities at $r=0$ and $r=a$. 
 
The scalar field is a real  function for any spacetime dimension provided $r$ and $a$ have the same sign. We choose  both of them positive without loss of generality. Due to the curvature singularity at $r=a$, the origin is set there, so that the radial coordinate ranges as $r>a$. Thus,  the lapse $f(r)$ and  scalar field $\Phi(r)$ are regular functions everywhere.

It is easy to verify that the lapse $f(r)$ in \eqref{f} is a monotonically increasing function and has a single  non-zero positive root only if $\alpha >0$. This root is given by
\begin{equation}
r_+=h \,a, \quad \mbox{with} \quad h=\left(1-\exp\left(-\frac{1}{\alpha l^2}\right)\right)^{-\frac{1}{n-1}}.
\end{equation}
Note that $h$ is fixed in terms of the parameters of the action and the spacetime dimension. For $\alpha >0$,  we have $h>1$ and therefore $r=r_+= h a$ corresponds to an event horizon which hides a curvature singularity at $r=a$.  As we are interested in black hole solutions, hereafter we consider $\alpha >0$ in the potential \eqref{potential}.

Expanding the potential \eqref{potential} around $\Phi=0$,
\begin{equation}
V(\Phi)\xrightarrow[\Phi\rightarrow 0]{}-\frac{(n-1)(n-2)}{2\kappa l^2}-\frac{(n-1)^3\alpha\kappa}{64n}\Phi^4+\mathcal{O}(\Phi^6),
\end{equation}
we observe that the leading term of the series corresponds to an $n-$dimensional negative cosmological constant, which is written in terms of the AdS radius $l$. Moreover, the next subleading term is proportional to $\Phi^4$. Thus, the potential does not contain a mass term. However, the non-minimal term  
$\xi R $ appearing in \eqref{KG} provides such a massive term. This can be seen as follows. The scalar field tends to zero for large $r$, and the metric locally approaches the AdS spacetime  
$$f(r) \xrightarrow[r \to \infty]{} \frac{r^2}{l^2}-\alpha \frac{a^{n-1}}{r^{n-3}}+\mathcal{O}(r^{-2n+4}). $$
Consequently, the Ricci scalar approaches to $R \to -n(n-1) l^{-2}$, so that at infinity the mass  term is
\begin{equation}
\xi R = -\frac{(n-1)^2}{4 l^2},
\end{equation}
which exactly matches the Breitenlohner-Freedman mass bound in $n$ spacetime dimensions. This is in agreement with the effective mass found in \cite{Lu_hairy}.

\section{Rotating hairy black holes}
\label{secrotating}

In this section a class of rotating hairy black holes is introduced. This class is built from the static black holes presented in the previous section. The starting point is to consider the set of  $(n-2)-$dimensional Ricci flat spaces having the  form  $\mathcal{M}_{n-2}= \mathbb{R} \times  \mathcal{M}_{n-3}$, where  $\mathcal{M}_{n-3}$ is a $(n-3)-$Ricci-flat space. Thus,  the line element of $\mathcal{M}_{n-2}$ can be split as
\begin{equation} \label{split}
d\sigma^{2}=d\phi^2+d \Sigma^2,
\end{equation}
where $d \Sigma^2$ denotes the line element of $\mathcal{M}_{n-3}$, which is independent of the  $r,\phi$ coordinates.

The second and crucial step is to consider the following improper gauge transformation:
\begin{equation} \label{coc}
t \to \frac{1}{\sqrt{1-\omega^2}}\left(t-l \omega \phi \right), \qquad
\phi \to \frac{1}{\sqrt{1-\omega^2}}\left(\phi-\frac{\omega}{l}t \right).
\end{equation}
This transformation is a boost parametrized by $\omega^2 <1$ in the $t-\phi$ plane. However, this boost is not a permitted global coordinate transformation, because it changes an exact 1-form into a closed but not exact 1-form, as was shown in \cite{Stachel}. Therefore, \eqref{coc} only preserves the local geometry, but not the global one. In consequence, the resulting spacetime is globally stationary but locally static. 

Applying \eqref{coc} to the static metric \eqref{staticsolution}  with a base manifold of the form \eqref{split}, we obtain the stationary axisymmetric line element
\begin{equation} \label{stationary}
ds^2=-\frac{  l^2 f(r)-r^2 \omega^2}{l^2 (1-\omega^2) }dt^2+\frac{2  \omega \left(l^2 f(r)-r^2\right)}{l (1-\omega^2) }dt d\phi +\frac{ r^2-l^2  \omega^2 f(r)}{1-\omega^2 }d\phi^2 +\frac{dr^2 }{f(r)}+r^2 d\Sigma^2.
\end{equation}
The transformation \eqref{coc} has no effect on the scalar field, so that it remains with the same expression given in \eqref{scalar}.  The stationary line element \eqref{stationary} can be written in the canonical form,
\begin{equation} \label{ADMmetric}
ds^{2}=-N^2(r)f(r)dt^{2}+\frac{dr^{2}}{f(r)}+H(r)(d \phi+N^{\phi}(r) dt)^2+r^2 d\Sigma ^{2},  
\end{equation}
where
\begin{eqnarray}
N^2(r)&=&\frac{r^2  (1-\omega^2)}{r^2-l^2 \omega^2 f(r) }, \\
N^{\phi}(r)&=&-\frac{ r^2-l^2 f(r)}{r^2-l^2  \omega^2 f(r) }\frac{\omega}{l},\\
H(r)&=&\frac{ r^2-l^2 \omega^2 f(r) }{1-\omega^2 }.
\end{eqnarray}

Since $f(r)$ in \eqref{f} has a single root for $\alpha >0$, the above stationary line element describes a rotating black hole with a single horizon located at $r= h a$, dressed with the scalar field \eqref{scalar}. The horizon surrounds the curvature singularity, which remains at the same place $r=a$. The functions $N^2(r)$ and  $H(r)$ are positive while the shift $N^{\phi}(r)$ is negative everywhere. Nevertheless, all of them are  monotonically increasing functions outside the horizon. 

In the asymptotic region, $r\to \infty$, we have 
\begin{eqnarray}
N^2(r)&=&1+\mathcal{O}(r^{-n+1}), \\
N^{\phi}(r)&=&-\frac{ \alpha l \omega a^{n-1}}{(1- \omega^2) r^{n-1}}+\mathcal{O}(r^{-2n+2}),\\
H(r)&=&r^2 +\frac{ \alpha l^2 \omega^2 a^{n-1}}{(1- \omega^2) r^{n-3}}+\mathcal{O}(r^{-2n+4}).
\end{eqnarray}

In consequence,  the rotating black hole is an asymptotically locally AdS spacetime in $n$ dimensions.

\section{Mass and angular momentum } \label{seccharges}

In this section we follow the Regge-Teitelboim approach \cite{ReggeTeitelboim} to determine the mass and angular momentum of the class of rotating hairy black holes shown in the previous section. In general, the canonical generator of an asymptotic symmetry described by the vector
$\xi^{\mu}=(\xi^{\perp},\xi^{i})$ is given by a linear combination of the Hamiltonian constraints
$\mathcal{H}_{\perp}, \mathcal{H}_{i}$, which is supplemented with a surface term $Q[\xi^{\mu}]$. Then, the generator reads
\begin{equation}\label{generator}
H[\xi^{\mu}]=\int d^{n-1} x \left(  \xi^{\perp}
\mathcal{H}_{\perp}+\xi^{i}\mathcal{H}_{i}\right) +Q[\xi^{\mu}].
\end{equation}
In order to ensure a well-defined generator, the term $Q[\xi^{\mu}]$ must cancel out the surface terms arising from the variation of the generator with respect to the canonical variables \cite{ReggeTeitelboim}. 

Note that the Hamiltonian constraints vanish for a specific solution. Thus,  the generator (\ref{generator}) reduces to the surface term $Q[\xi^{\mu}]$. In consequence, the conserved charges are just given by this surface term.  Since $Q[\xi^{\mu}]$ is 
defined at the boundary, the computation of the charges requires only the asymptotic behavior of the canonical variables and symmetries. 

We consider a minisuperspace containing the set of stationary metric of the form \eqref{ADMmetric} and a scalar field just depending on the radial coordinate. In this way, the expressions for the constraints are
\begin{align}
\mathcal{H_{\bot}}=&-\sqrt{\frac{H \gamma}{f}}r^{n-3}\left[(1-\kappa\xi\Phi^2)\left(\frac{^{(n-1)}R}{2\kappa}\right)-\frac{1}{2} f r^{n-3}\Phi^{\prime 2}-V\right] \nonumber \\&+\sqrt{\frac{f}{H\gamma}}\left(\frac{4\kappa}{1-\kappa\xi\Phi^2}\right)\pi_{r\phi}\pi^{r\phi}-[\xi \sqrt{f H \gamma}r^{n-3}(\Phi^{2})^{\prime}]^{\prime},  \label{hperp}\\
\mathcal{H}_{\phi}=&-2{{\pi_{\phi}}^{r}}_{|r}, \label{Hphi}
 \end{align} 
where ${}^{\prime}$ stands for $d/d r$, ${}^{(n-1)}R$ is the Ricci scalar of the spatial section of the metric $\eqref{ADMmetric}$ and $\gamma$ is the determinant of  $d\Sigma^2$  in the same metric. The only nonvanishing canonical momentum of the gravitational field is given by
\begin{equation}
\pi_{\phi}^{\ r}=-\frac{H}{N}\sqrt{H \gamma}r^{n-3}\left(\frac{1-\kappa\xi\Phi^2}{4\kappa}\right) (N^{\phi})^\prime. 
\end{equation}

Using \eqref{hperp} and \eqref{Hphi} we obtain the variation of the charge
\begin{align} 
\delta Q[\xi^{\mu}]&=  2 \pi \Sigma \lim_{r \to \infty} \left\{\sqrt{\frac{H\gamma}{f}}\left[\left(-\frac{f}{H}(\delta H'-\frac{H'}{2H}\delta H)-(\frac{H'}{2H}+\frac{n-3}{r})\delta f\right)\left(\frac{1-\kappa\xi\Phi^2}{2\kappa}\right)\xi^{\bot} \right.\right. \nonumber \\&+\left.\left. \frac{f}{H}  \left[ \left(\frac{1-\kappa\xi\Phi^2}{2\kappa}\right)\xi^{\bot}\right]'\delta H +f\left[ \xi^{\bot}(\xi(\delta\Phi^2)'-\delta\Phi \Phi')-\xi^{\bot}_{,r}\xi\delta\Phi^2 \right] \right]+2\xi^{\phi}\delta\pi_{\phi}^{\ r}\right\}, \label{dq}
\end{align}
where the integration over the coordinate $\phi$ and the base space $\mathcal{M}_{n-3}$, whose volume is denoted by $\Sigma$, was done. Additionally, the components of the asymptotic symmetries are given in terms of the spacetime components  $ \partial_t$ and  $ \partial_\phi$ as follows,
\begin{align}
\xi^{\bot}&=N\sqrt{f}  \partial_t,\\
\xi^{\phi}&=  \partial_\phi+N^{\phi}  \partial_t.
\end{align}

Now, we focus on the rotating black hole solution introduced in the previous section. The asymptotic variation of the fields is of the form
\begin{eqnarray}
\delta f(r)&=&\alpha(1-n)a^{n-2}\frac{\delta a}{r^{n-3}}+\mathcal{O}(r^{-2n+4})\ ,\\
\delta H(r)&=&\left(2 a \delta\omega-\omega(1-n)(1-\omega^2)\delta a\right)\frac{\alpha a^{n-2}\omega l^2}{(1-\omega^2)^2 r^{n-3}}+\mathcal{O}(r^{-2n+4})\ ,\\
\delta \Phi(r)^2&=&4 n a^{n-2}\frac{\delta a}{\kappa r^{n-1}}\ ,\\
\delta\pi_{\phi}^{\ r}&=&\frac{\alpha l (1-n)}{4\kappa (1-\omega^2)}\delta{(\omega a^{n-1})}+\mathcal{O}(r^{1-n}). \label{dpi}
\end{eqnarray}

The black holes considered here have two symmetries. The time-translation symmetry defined by the Killing vector $\partial_t$, whose conserved charge is the mass $M$,  and the rotational symmetry generated by the Killing vector $\partial_\phi$, whose charge is the angular momentum $J$. 

Using the asymptotic form of fields and their variations, we obtain from \eqref{dq} integrable expressions for the mass $M= Q(\partial_t)$ and the angular momentum $ J=-Q(\partial_\phi)$, which are given by
\begin{eqnarray}
M&=&\frac{\pi}{\kappa}\Sigma\alpha a^{n-1}\left(\frac{\omega^2+n-2}{1-\omega^2}\right), \label{mass}\\
J&=&\frac{\pi}{\kappa}\Sigma\alpha a^{n-1} l \omega \left(\frac{n-1}{1-\omega^2}\right). \label{J}
\end{eqnarray}
up to additive fixed constants, which have been set to vanish in order to have a massless and static background in absence of the scalar field ($a=0$). Thus, the background configuration takes the form,
\begin{equation} \label{bg}
ds^2=-\frac{r^2}{l^2}dt^2+\frac{l^2}{r^2}d r^2+r^2 d\sigma^{2}, \quad \Phi=0.
\end{equation}

Naturally, the angular momentum vanishes for $\omega=0$ and the mass is non-negative because $\omega^2 <1$. Moreover, from \eqref{mass}-\eqref{J} it is possible to show that the angular momentum is bounded from above by the mass,  $M >| J / l|$. In fact, this bound guarantees the existence of an event horizon as can be seen from the expression of its radius in terms of $M$ and $J$,
\begin{equation}
r_+=h \left[\frac{\kappa}{2\pi\alpha l \Sigma}\frac{\sqrt{M^2 l^2(n-1)^2-4J^2(n-2)}-M l (n-3)}{n-2}    \right]^{\frac{1}{n-1}}.
\end{equation}
Additionally, the angular velocity reads
\begin{equation}
\omega=\frac{Ml(n-1)-\sqrt{M^2 l^2(n-1)^2-4J^2(n-2)}}{2J},
\end{equation}
in agreement with the condition $\omega^2 <1$ by virtue of the bound $M >| J / l|$.

\section{Thermodynamics} \label{thermo}

In this section the thermodynamical description of the rotating hairy black holes is presented and their local and global stability is analyzed. 

The temperature can be determined by means of the surface gravity $k$ at the horizon. This is achieved by considering the Killing vector $\xi= \partial_t- \Omega \partial_\phi$, where $\Omega=-N^{\phi}(r_+)=\omega/l$ is the angular velocity at the horizon. Thus, the temperature is given by
\begin{equation} \label{temperatura}
T=\frac{k}{2 \pi}=\frac{\alpha  h (n-1) a \sqrt{1-\omega^2}}{4 \pi  \left(h^{n-1}-1\right)}.
\end{equation}
Due to the nonminimal coupling, the standard  Bekenstein-Hawking entropy acquires an extra factor \cite{Visser, Ashtekar} and reads
\begin{equation}
S=\left(1-\kappa \xi \Phi(r_+)^{2}\right) \frac{ 2 \pi A_+}{\kappa},
\end{equation}
where $A_+$ is the area of the horizon. Then, for $r_+ = h a$, the entropy is 
\begin{equation}
S=\frac{ 4 \pi^2}{\kappa} \frac{ a^{n-2}\Sigma }{\sqrt{1-\omega^2 }} \left( \frac{h^{n-1}-1}{h}\right).
\label{entropia}
\end{equation}
Since $\omega^2 <1$ and $h>1$ the temperature and entropy are real and positive quantities.
Furthermore, the first law $dM=T dS+\Omega d J$ is verified and the Smarr formula takes the form
\begin{equation} \label{Smarr}
M= \frac{n-2}{n-1} T S +\Omega J.
\end{equation}

\subsection{Local stability}

The thermodynamical stability of a system requires to analyze the response of the system under small perturbations of its thermodynamical variables around the equilibrium. There are some equivalent criteria for studying stability, as for example, the sign of second derivative of the entropy as well as the energy, or any of its associated Legendre transforms. In consequence, the preference on one free energy respect to another is clearly a matter of convenience, but the stability requirements depend on the kind of themodynamical ensemble considered. In this section the grand canonical ensemble is chosen, where the temperature $T$ and angular velocity $\Omega$ are fixed. In this ensemble, the thermodynamical stability analysis can be performed by using the Gibbs free energy $G(T,\Omega)$. The local stability criteria demand to analyze the concavity of this function (see for instance \cite{Callen}), which implies the following stability conditions: 
\begin{equation} \label{stacon}
\mbox{(i)} \; \;\frac{\partial^2 G}{\partial T^2} \leq 0, \quad \mbox{(ii)}\; \; \frac{\partial^2 G}{\partial \Omega^2} \leq 0, \quad \mbox{(iii)}  \; \;\frac{\partial^2 G}{\partial T^2}\frac{\partial^2 G}{\partial \Omega^2}-\left(\frac{\partial^2 G}{\partial T\partial \Omega }\right)^2 \ge 0.
\end{equation}
By using the Smarr relation \eqref{Smarr},  the Gibbs free energy 
$G=G(T, \Omega)= M-T S -\Omega J$,
reduces to
\begin{equation} \label{Gibbs2}
G= -\frac{T S}{n-1}  = -\frac{\pi}{\kappa} \Sigma \alpha   a^{n-1}.
\end{equation}
Now, we need to replace $a$ in terms of $T$ and $\Omega$ to get   
\begin{equation} \label{Gibbs3}
G(T, \Omega)=  -\frac{\pi \Sigma \alpha }{\kappa} \left[\frac{4 \pi  \left(h^{n-1}-1\right)}{\alpha  h (n-1) } \frac{T}{\sqrt{1-\Omega^2 l^2}}\right]^{n-1}.
\end{equation}
Considering the stability criteria one gets the following expressions,
\begin{align}
\frac{\partial^2 G}{\partial T^2}  = &  -\frac{  4^{n-1}\alpha (n-2) (n-1) \pi ^n \Sigma  \left(h^{n-1}-1\right)^{n-1} T^{n-3}}{\kappa \left(\alpha  h (n-1) \sqrt{1- \Omega^2 l^2}\right)^{n-1} } \leq 0,\\
\frac{\partial^2 G}{\partial \Omega^2} = & -\frac{ 4^{n-1}\alpha  l^2 (n-1) \pi ^n \Sigma  \left(n \Omega^2 l^2 +1\right) \left(h^{n-1}-1\right)^{n-1} T^{n-1} }{\kappa  \left(1- \Omega^2 l^2\right)^2 \left(\alpha  h (n-1) \sqrt{1- \Omega^2 l^2}\right)^{n-1} } \leq 0,
\end{align}
and
\begin{align}
\frac{\partial^2 G}{\partial T^2}\frac{\partial^2 G}{\partial \Omega^2}-\left(\frac{\partial^2 G}{\partial T\partial \Omega }\right)^2 = & \frac{4^{2n-2}\alpha ^4 h^2 l^2  (n-1)^4 \pi ^{2 n} \Sigma ^2  \left(h^{n-1}-1\right)^{2n-2} T^{2n-4}  }{\kappa ^2  \left(\alpha  h (n-1) \sqrt{1- \Omega^2 l^2}\right)^{2n}} \nonumber \\ \times& \left( 1+ \frac{n-3}{1- \Omega^2 l^2}\right) \ge 0 .
\end{align}
Therefore, the rotating hairy black holes are thermodynamically locally stable in any dimension $n \ge 3$.

An alternative necessary condition for local thermal stability comes from the specific heat at fixed angular velocity,
\begin{equation}
C_{\Omega}=\left(\frac{\partial M}{\partial T}\right)_{\Omega}=4\pi^2\Sigma r_+^{n-2}\frac{(1-h^{1-n}) (\omega^2+n-2)}{\kappa(1-\omega^2)^{3/2}},
\end{equation}
which is always positive. This is in agreement with the concavity of the Gibbs free energy, and in consequence, the rotating hairy black hole always attains equilibrium with a heat bath.

\subsection{Global stability}
Note that for the hairy black hole it is not possible to switch off the scalar field and, at the same time, keep the mass fixed. In fact, in the limit $\Phi\to 0$ the black hole metric tends to the background one given in \eqref{bg}.  Therefore, for a fixed mass there are two branches of different black hole solutions with the same topology, the hairy black hole, and the vacuum solution with $\Phi=0$, which corresponds to the $n$ dimensional generalization of Lemos black string \cite{Lemos01}, whose metric reads
\begin{equation}
ds_0^2=-\left(\frac{\rho^2}{l^2}-\frac{b l^{n-3}}{\rho^{n-3}}\right)dt^2+\left(\frac{\rho^2}{l^2}-\frac{b l^{n-3}}{\rho^{n-3}}\right)^{-1}d\rho^2+\rho^2 d\sigma^{2},
\end{equation}
where $b >0$ is an integration constant and $d\sigma^{2}$  is the line element shown in \eqref{staticsolution}.

It is possible to add angular momentum to this solution proceeding in the same way as in Sec. \ref{secrotating}.  Thus, we obtain
the line element \eqref{stationary}, but now the radial coordinate is denoted by $\rho$ and the function $f$ is replaced by
\begin{equation}
f(\rho)=\frac{\rho^2}{l^2}-\frac{b l^{n-3}}{\rho^{n-3}},
\end{equation}
where the boost parameter is now denoted by $\omega_0$.

Global stability can be analyzed by comparing the free energy of the hairy black hole and the vacuum solution. To achieve this we use the grand canonical ensemble by setting both configurations at the same temperature $T$ and angular velocity $\Omega$.

The temperature, angular velocity and entropy for the vacuum solution are
\begin{equation}
T_0=\frac{(n-1)b^{1/(n-1)}\sqrt{1-\omega_0^2}}{4\pi l}, \quad \Omega_0=\frac{\omega_0}{l}, \quad 
S_0=\frac{ 4 \pi^2}{\kappa} \frac{ \Sigma b^{(n-2)/(n-1)}l^{n-2}}{\sqrt{1-\omega_0^2 }},
\end{equation}
respectively. Furthermore, using  the variation of the charges \eqref{dq} and the asymptotic form of the rotating vacuum solution, the mass and angular momentum\footnote{Note that this vacuum solution verifies the same Smarr relation shown previously for the hairy black hole in \eqref{Smarr}. } are determined as
\begin{align}
M_0&=\frac{\pi}{\kappa}\Sigma b l^{n-3}\left(\frac{\omega_0^2+n-2}{1-\omega_0^2}\right),\\
J_0&=\frac{\pi}{\kappa}\Sigma b l^{n-2}  \omega_0 \left(\frac{n-1}{1-\omega_0^2}\right).
\end{align}

Therefore, the Gibbs free energy of the vacuum solution is 
\begin{equation}
G_0= -\frac{\pi \Sigma b l ^{n-3}}{\kappa}.
\end{equation}
Then, to compare both black holes in the grand canonical ensemble,  their temperatures and angular velocities must coincide yielding,
\begin{equation}
a=\frac{(h^n-h)b^{1/(n-1)}}{h^2 \alpha l}, \quad \omega=\omega_0,
\end{equation}
respectively.

Considering the above matching conditions,  the difference between the free energies of the hairy and vacuum black holes gives
\begin{equation}
\Delta G=G-G_0=\frac{\pi\Sigma l^{n-3} b(T_0,\Omega_0)}{\kappa}g\left(\frac{1}{\alpha l^2}\right),
\end{equation}
where 
\begin{equation}
g(x)= 1-\left(\frac{x}{e^x-1}\right)^{n-2} e^{-x}
\end{equation}
is a positive monotonically increasing function for $x>0$. Then, since $b(T_0,\Omega_0)>0$, we conclude that $\Delta G >0$. Therefore the vacuum solution is the configuration thermodynamically preferred over the hairy one.

\section{Concluding remarks} \label{remarks}

A class of exact rotating black hole solutions of gravity nonminimally coupled to a self-interacting scalar field in arbitrary dimensions has been presented. These geometries are asymptotically locally AdS spacetimes and possess an event horizon with a Ricci-flat geometry that covers the curvature singularity at the origin. The scalar field is real and regular everywhere.

The existence of black holes is ensured for a positive self-interaction coupling constant. In this case the self-interaction potential $V(\Phi)$ is negative, concave and unbounded from below. It is interesting to note that the mass term for scalar perturbations around the equilibrium point $\Phi=0$, is provided by the nonminimal coupling term, and not by the potential as it is usually expected. The non-minimal parameter possesses a precise dependence on the dimension in such a way that the effective mass term saturates the Breitenlohner-Freedman bound for arbitrary dimension, which ensures the perturbative stability of the global AdS spacetime under scalar perturbations.

The rotating black hole solution is obtained by applying an improper coordinate transformation to the static one. In consequence, although both spacetimes are locally equivalent, they are globally different. In this work, a single boost in the plane $t-\phi$ was considered. However, it is possible to add angular momentum in other planes by consecutive boosts.

The mass and angular momentum were computed using the Regge-Teitelboim procedure. It is found that the mass is bounded from below by the angular momentum as $M>|J/l|$, being consistent with the existence of an event horizon.

The thermodynamical analysis was carried out in the grand canonical ensemble.  We found that the first law is satisfied and a Smarr formula is exhibited. Due to the nonminimal coupling of the scalar field with the curvature, the standard area law for the entropy is modified by an extra factor, which depends on the value of the scalar field at the horizon. In our case, this factor is always positive ensuring the positivity of the entropy for the rotating hairy black hole. The Gibbs energy of the black hole is a concave function of the temperature and angular velocity and, in consequence, the rotating hairy black holes are thermodynamically locally stable in any dimension $n\geq 3$. However, in the global stability analysis, it was found that the vacuum spacetime is always favored over the hairy black hole. In other words, the hairy black hole is likely to decay into the vacuum black hole for any temperature.

\subsection*{Acknowledgements}
We thank Ernesto Frodden for his participation in the early stages of this article and Ricardo Troncoso for useful discussions. This work has been partially funded by the Fondecyt grant 1161311. The Centro de Estudios Cient\'{\i}ficos (CECs) is funded by the Chilean Government through the Centers of Excellence Base Financing Program of Conicyt.

\newpage
\bibliographystyle{science}
\bibliography{CMCE02}

\end{document}